\documentclass[10pt,twocolumn,letterpaper]{article}

\usepackage{cvpr}              

\usepackage{booktabs}
\usepackage{multirow}
\usepackage{makecell}
\usepackage{multirow}
\usepackage{booktabs}
\usepackage{graphicx} 
\usepackage{siunitx}

\definecolor{cvprblue}{rgb}{0.21,0.49,0.74}
\usepackage[pagebackref,breaklinks,colorlinks,allcolors=cvprblue]{hyperref}

\def\paperID{MEIS-9} 
\def\confName{CVPR}
\def\confYear{2026}

\title{Multi-Agent Video Prediction: Self-Correcting Conditional Frames for Dynamic Scene Forecasting}

\author{
Qixin Zhang \quad Ajay Kumar Gurumadaiah \quad Zhi-Li Zhang\\
University of Minnesota\\
{\tt\small \{zhan8548, gurum021, zhang089\}@umn.edu}
}

\begin{document}
\maketitle
\begin{abstract}
Transmission latency significantly degrades user quality of experience in real-time interactive perception systems. In remote driving, maintaining reliable visual feedback is critical for safe operation under dynamic network variability. Although video prediction offers a promising approach to compensate for short-term transmission delays and approximate near-zero-latency streaming, prediction-only methods remain vulnerable in highly dynamic scenes, especially when newly emerged objects appear during uplink outages. To address these challenges, we propose a multi-agent video prediction framework that combines continuous edge-side video prediction with lightweight mask-guided conditional frame reconditioning. The framework consists of three role-specialized agents: a continuous prediction agent for low-latency visual continuity, a vehicle-side trigger agent for detecting newly appeared objects, and a conditional reconditioning agent that repairs the predictor conditioning state using sparse mask guidance. This design enables semantic recovery of exogenous scene changes without requiring full-frame retransmission. We validate the proposed framework through extensive experiments on benchmark video data under realistic 5G communication traces. Results show that our method improves semantic recovery of novel objects while preserving perceptual quality and practical runtime efficiency under network-induced disruptions.
\end{abstract}

\section{Introduction}
\label{sec:intro}
Real-time visual communication has become a core component of modern interactive systems, enabled by advances in wireless networking, cloud infrastructure, and edge computing \cite{sato2023compensation}. Applications such as remote collaboration, interactive streaming, surveillance, and online gaming rely on continuous video transmission to support real-time perception and interaction, making low end-to-end latency critical for smooth and reliable operation. This requirement becomes even more important in safety-critical and human-in-the-loop scenarios, including telesurgery, remote robotic manipulation, and remote vehicle operation, where operators depend on timely visual feedback to make accurate control decisions \cite{hatano2023evaluation, yu2022remote}. However, delays introduced by network transmission, video encoding, and distributed processing can degrade situational awareness and response time, making latency mitigation a key challenge in teleoperation systems ~\cite{wang2021inferring, mody2014ultra}.

Recent advances in deep generative models, including diffusion-based and autoregressive architectures, have significantly improved the fidelity and temporal coherence of predicted video frames. Models such as MCVD ~\cite{voleti2022mcvd} and ExtDM ~\cite{zhang2024extdm} demonstrate the ability to synthesize plausible future frames conditioned on past observations. However, their effectiveness in real-world teleoperation scenarios remains limited due to the highly dynamic and interactive nature of environments such as road traffic. Small prediction errors can accumulate over time, leading to drift, inconsistent object motion, or unrealistic scene evolution, particularly when the model lacks a deeper understanding of the underlying scene dynamics

\begin{figure*}[t]
    \centering
    \includegraphics[width=0.95\linewidth]{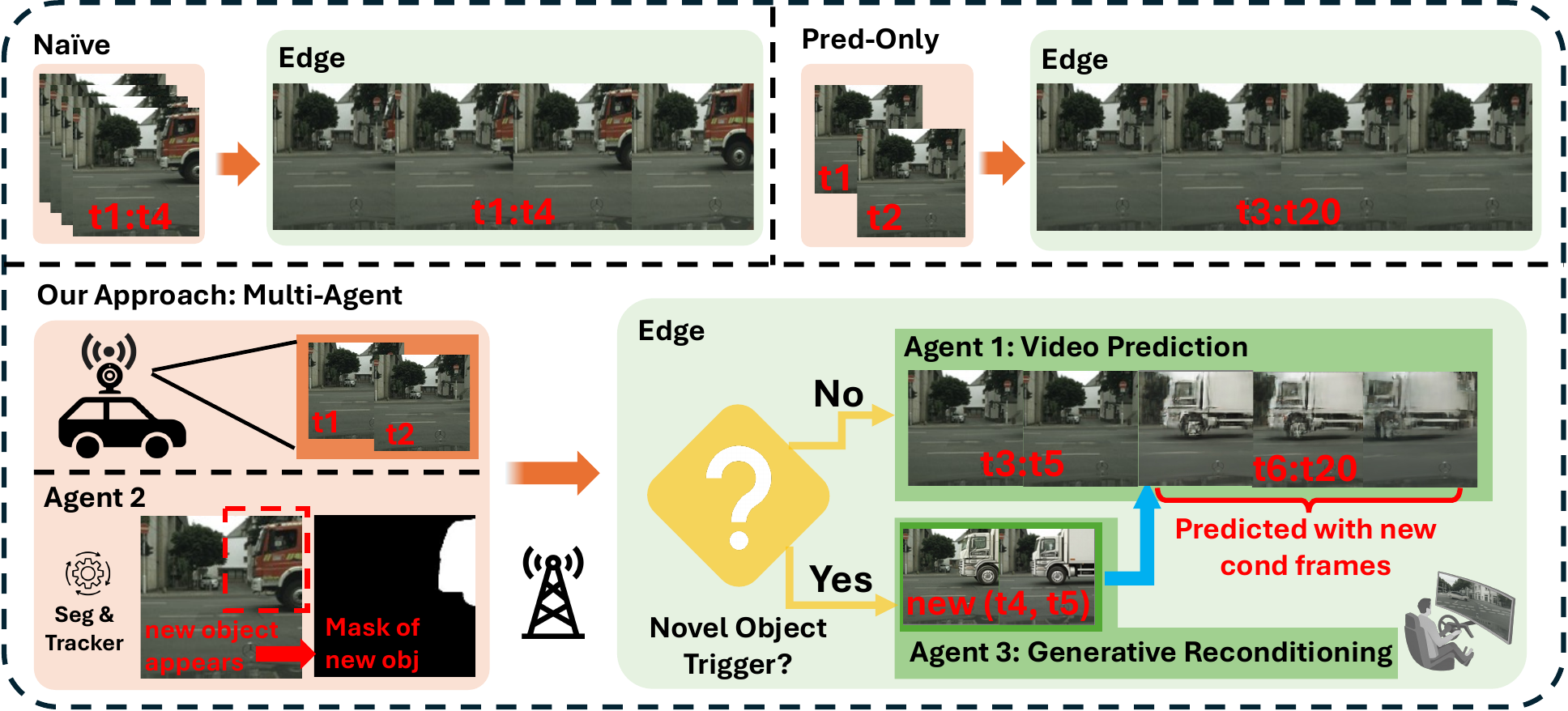}
    \caption{Architecture of Our Multi-Agent Approach.}
    
    \label{fig:architecture}
    \vspace{-1.5em}
\end{figure*}

To address these limitations, we propose a Multi-Agentic Video Prediction framework that introduces collaborative agents to improve prediction robustness in dynamic environments. Instead of relying on a single predictive model, our framework employs specialized agents that maintain visual continuity and restore missing scene semantics during communication gaps. A prediction agent first generates future frames, while auxiliary agents analyze scene dynamics and generate refined conditional frames to correct accumulated drift or noise. This collaborative mechanism enables the system to maintain consistency with the underlying scene structure and motion, leading to more reliable video prediction under real-world teleoperation conditions. Figure~\ref{fig:architecture} provides a high-level overview of the proposed multi-agent framework.

Motivated by these challenges, we propose a Multi-Agentic Video Prediction framework that integrates continuous prediction, novel-object triggering, and conditional reconditioning to repair the predictor conditioning state. This collaborative mechanism improves temporal consistency and robustness in dynamic environments. The key contributions of this work are summarized as follows:

\begin{itemize}
   \item We propose a multi-agent reconditioning framework for predicted video streaming that combines continuous edge prediction with lightweight mask-guided inpainting to inject newly emerged objects into the conditioning state.
    \item We identify and formalize a structural failure case in prediction-only streaming under open-world communication gaps, termed Novel-Object Blindness (NOB), which motivates the need for reconditioning.
    \item We provide qualitative and quantitative evidence under realistic 5G uplink traces that the proposed multi-agent framework improves semantic recovery while maintaining perceptual quality and practical runtime overhead.
\end{itemize}

\section{Related Work}
\label{sec:related_work}

\subsection{Multi-Agent Coordination in Embodied Systems}
Embodied AI research has established interactive environments and benchmark suites for grounding perception, planning, and action under partial observability, including Habitat, ALFRED, and TEACh \cite{savva2019habitat, shridhar2020alfred, padmakumar2021teach}. Recent agentic frameworks further explore role specialization and inter-agent communication for complex tasks, including CAMEL, AutoGen, and Voyager \cite{li2023camel, wu2023autogen, wang2023voyager}. These studies show that decomposing a task into specialized agent roles can improve adaptability and robustness in open-ended settings.

Most of these systems, however, optimize for task-level outcomes such as instruction completion, dialogue quality, or long-horizon exploration \cite{padmakumar2021teach, li2023camel, wu2023autogen, wang2023voyager}. Their coordination primitives are typically designed for planning and decision-making over symbolic or text-centric state representations, rather than for preserving semantic fidelity in a continuously generated visual stream \cite{li2023camel, wu2023autogen}. In communication-constrained teleoperation, the central requirement is different: agents must coordinate online to maintain both temporal continuity and object-level scene correctness under intermittent observations \cite{hirche2004packet, kizilkaya20235g}.

However, existing multi-agent literature is primarily centered on task completion, dialogue planning, or long-horizon policy execution. It does not directly address semantic reliability in predictive video streams under communication outages. Our work targets this gap by framing communication-gap recovery as role-specialized coordination between a continuity agent and a state-correction agent.

\subsection{Video Prediction Under Partial Observability}
Autoregressive video prediction methods extrapolate future frames from a fixed history window. Early recurrent models such as ConvLSTM and PredNet established temporal latent-state propagation \cite{shi2015convolutional, lotter2017deep}, and subsequent work improved motion-content modeling for longer horizons \cite{villegas2017decomposing}. Diffusion-based video models, including Video Diffusion Models and MCVD, have further improved perceptual quality in conditional generation \cite{ho2022video, voleti2022mcvd}.

Frame interpolation methods such as Super-SloMo and RIFE produce high-quality in-between frames when both temporal anchors are available \cite{jiang2018super, huang2020rife}; however, they are not applicable during real-time transmission outages where future observations are unavailable. More broadly, high-fidelity diffusion models \cite{ho2020denoising, saharia2022photorealistic, blattmann2023align} prioritize perceptual realism but do not explicitly guarantee object-level semantic correctness under exogenous scene changes.

Under communication gaps, this distinction is critical. A method can produce visually plausible and temporally smooth predictions while still failing to update scene state when novel entities emerge outside the conditioning history. This failure is structural rather than cosmetic: the issue is not merely reconstruction error, but the absence of an explicit mechanism for exogenous evidence injection into the predictive state. As a result, standard quality metrics alone may understate safety-relevant semantic omissions in teleoperation contexts \cite{hirche2004packet, kizilkaya20235g}.

Overall, prior video modeling methods emphasize realism or temporal smoothness under closed-history conditioning. They do not provide an explicit mechanism for injecting newly observed exogenous evidence into an ongoing predictive stream during communication gaps.

\subsection{Communication-Constrained Teleoperation and Streaming}
Prior teleoperation and streaming research has extensively studied low-latency transmission and network robustness, including DASH-style adaptation, edge-assisted delivery, and delay/packet-loss analysis \cite{stockhammer2011dash, bentaleb2018survey, mach2017mobile, shi2016edge, hirche2004packet, kizilkaya20235g}. These methods are essential for system-level reliability, but they largely treat visual content as passively transmitted data.

In contrast, our focus is semantic reliability under communication constraints: when outages occur, the stream must remain both temporally continuous and semantically correct. This requires coordination between predictive generation and targeted state correction, rather than transmission adaptation alone.

\section{Problem Statement}
\label{sec:pro_state}

Video prediction plays a critical role in latency-sensitive teleoperation systems where future frames must be estimated when new visual observations are delayed or unavailable. However, existing video prediction methods are primarily designed for offline forecasting settings and do not explicitly address the challenges introduced by communication latency, dynamic environments, and incomplete observations. To better understand these limitations, we formulate the following key research questions that guide the design of our approach.

\noindent \textit{Q1. Edge Extrapolation and the Single-Agent Limitation:} In latency-compensated teleoperation systems, video prediction often relies on the most recently received frame when network delays prevent timely updates. Existing methods typically use a single predictive agent that extrapolates scene dynamics from this conditional frame using learned motion representations such as optical flow or latent motion features. While effective for short-term prediction, this approach becomes unreliable in dynamic environments. As predictions recursively depend on previously generated frames, errors accumulate over time, leading to temporal drift and inconsistent scene evolution.

\noindent \textit{Q2. Novel-Object Blindness under Communication Gaps:} Another challenge arises during communication gaps when the teleoperation system fails to receive new frames from the remote environment. During these interruptions, prediction models generate frames based only on previously observed data, making them unaware of new objects or events that may appear in the scene. For example, in remote driving scenarios, new vehicles or pedestrians may enter the scene while communication is disrupted. As a result, the predicted frames may remain visually plausible but fail to reflect the true evolving state of the environment. We term this failure mode \emph{Novel-Object Blindness} (NOB), where prediction-only rollout fails to recover newly appeared objects that are absent from the current conditioning history.

\begin{figure}[t]
  \centering
  \includegraphics[width=0.95\linewidth]{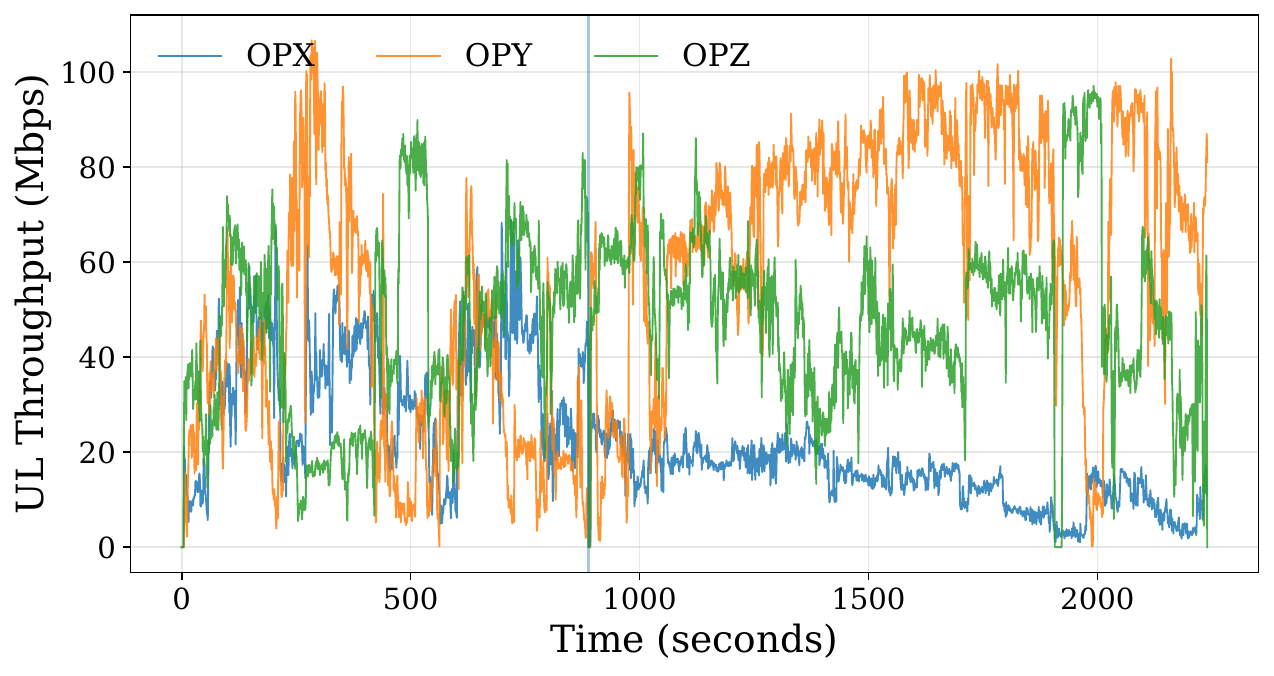}

   \caption{Uplink throughput over time in a 5G network for three operators in the U.S.}
   \label{fig:uplink_analysis}
\end{figure}

\noindent \textit{Q3. Uplink Instability and Perceptual Discontinuity in Teleoperation:}
In real-world teleoperation systems, video streams are transmitted over wireless networks such as 5G, where uplink bandwidth and latency can fluctuate due to channel variability and network congestion, as shown in Figure~\ref{fig:uplink_analysis}. These fluctuations can cause irregular frame delivery or temporary interruptions. During such disruptions, prediction models must generate frames without synchronized updates from the remote environment. When delayed frames arrive, discrepancies between predicted and actual scenes can produce abrupt visual changes, leading to perceptual discontinuities that degrade operator situational awareness.

Motivated by the above research questions, we define the problem of robust video prediction under communication latency and dynamic multi-agent environments. Let $X_{t-k:t} = \{X_{t-k}, \dots, X_t\}$ denote a sequence of k+1 observed frames up to time t. The objective is to learn a prediction model that estimates future frames over a prediction horizon $T$  that estimates the future frame sequence. 

\begin{equation}
\hat{X}_{t+1:t+T} = f(X_{t-k:t})
\end{equation}

where $T$ denotes the prediction horizon and $\hat{X}_{t+1:t+T}$ represents the predicted future frames. Recent approaches, including autoregressive and diffusion-based models, generate future frames by conditioning on previously observed frames and iteratively predicting subsequent frames.
\begin{equation}
\hat{X}_{t+i} = f(X_{t-k:t}, \hat{X}_{t+1:t+i-1}), \quad i = 1, \dots, T
\end{equation}
While these models have demonstrated promising results for short-term prediction in relatively structured environments, their performance degrades in highly dynamic scenarios.

\noindent In real-world applications such as remote driving and robotic teleoperation, environments often contain multiple interacting agents (e.g., vehicles, pedestrians, cyclists) whose motions evolve unpredictably. When prediction models rely solely on previously predicted frames as conditioning inputs, small inaccuracies in motion or appearance can accumulate over time, leading to temporal drift, motion inconsistencies, and unrealistic scene evolution. This problem becomes more severe when prediction is used to compensate for communication latency, where predictions must remain stable and physically plausible for extended horizons.

\begin{equation}
\epsilon_{t+i} = X_{t+i} - \hat{X}_{t+i}
\end{equation}

Consider a teleoperation video stream with intermittent uplink frame delivery. Let $X_t$ denote the ground-truth frame at time $t$, and let $c_t$ denote the conditioning state available at the edge for future prediction. In a prediction-only system, when new uplink observations are missing, the edge predictor rolls out future frames autoregressively from stale conditioning inputs. Formally, the predicted stream is generated as
\begin{equation}
\hat{X}_{t+i} = f_{\text{pred}}(c_t,\hat{X}_{t+1:t+i-1}), \quad i=1,\dots,T .
\end{equation}
where $f_{\text{pred}}$ denotes the video prediction model and $T$ is the rollout horizon.

This formulation is effective for preserving short-term temporal continuity, but it assumes that all semantically relevant scene content is already represented in the conditioning state. In practice, this assumption fails in open-world teleoperation. If a previously unseen object enters the scene during a communication gap, that object is absent from the edge-side conditioning history and therefore cannot be inferred reliably by prediction alone. As a result, the predicted stream may remain temporally smooth while becoming semantically incomplete.

Our goal is therefore not only to maintain visual continuity during uplink outages, but also to restore missing scene semantics when such exogenous events occur. To do so, we seek a mechanism that (i) preserves the low-latency benefits of edge-side prediction, (ii) avoids the bandwidth cost of retransmitting full frames, and (iii) updates the predictor using sparse corrective evidence when a novel object appears. In our framework, this corrective evidence is provided as a lightweight object mask and is used to repair the predictor conditioning state before autoregressive rollout continues.

\begin{figure}[t]
  \centering
  \includegraphics[width=0.85\linewidth]{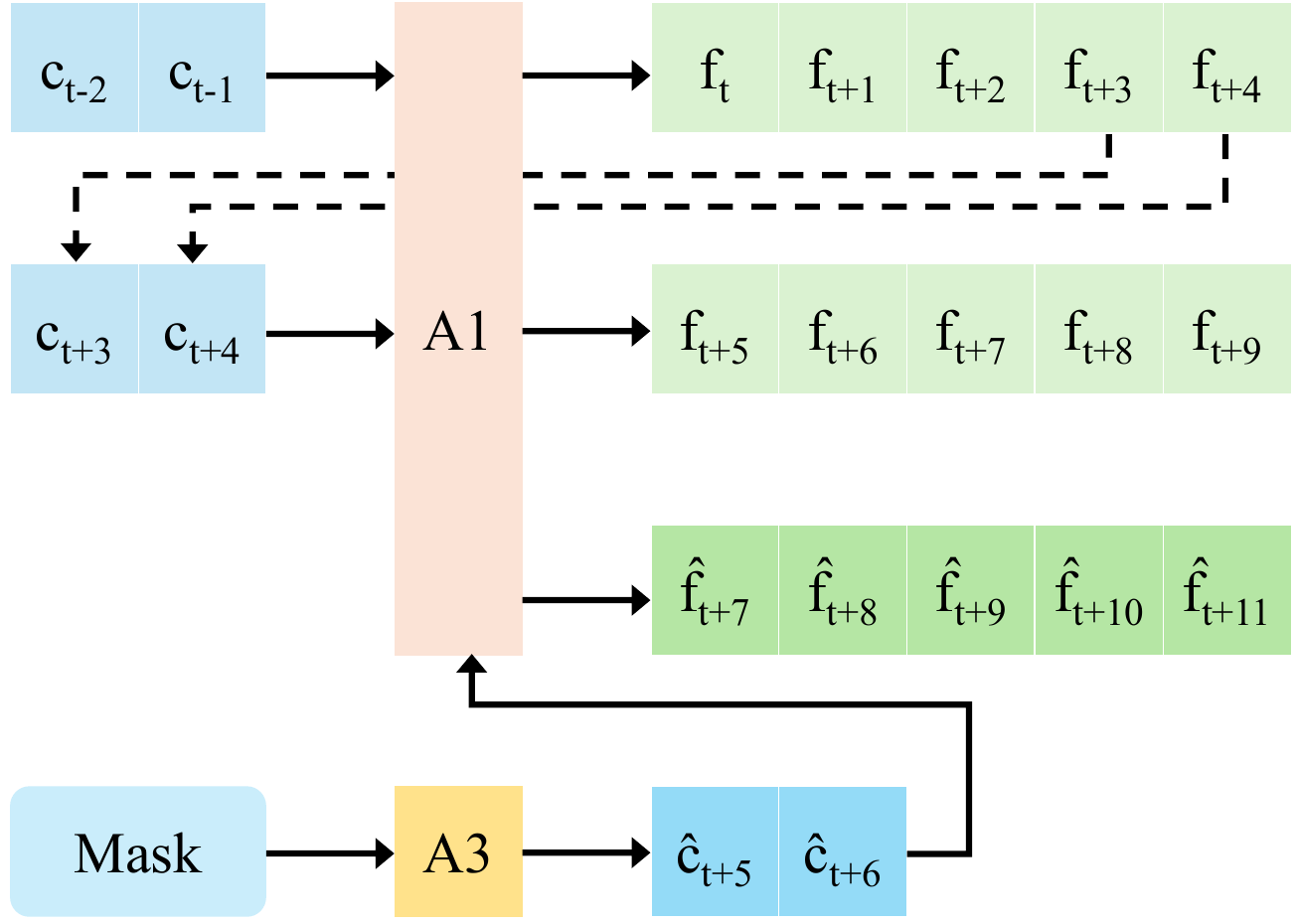}
   \caption{Multi-agent prediction and reconditioning pipeline on the edge. Notation: $c_t$ conditioning frames; $\hat{c}_t$ repaired conditioning frames; $f_t$ future frames predicted by $A1$; $\hat{f}_t$ reconditioned future frames.}
   \label{fig:flow}
\end{figure}

\section{Agentic Video Prediction Framework}
\label{method}
\subsection{System Overview}
We propose an agentic video prediction framework for maintaining continuous visual feedback under intermittent uplink transmission while correcting semantic omissions caused by prediction-only rollout. The framework consists of three role-specialized agents deployed across the vehicle and the edge: (i) a Continuous Prediction Agent (Agent 1) on the edge, (ii) a Novel-Object Trigger Agent (Agent 2) on the vehicle, and (iii) a Conditional Frame Reconditioning Agent (Agent 3) on the edge. Figure~\ref{fig:flow} illustrates the overall pipeline.

Agent 1 is responsible for temporal continuity. In our implementation, it is instantiated with MCVD, although the framework itself is predictor-agnostic. Agent 2 monitors the live onboard stream and detects newly appeared objects using segmentation and tracking; in our experiments, we use a YOLO-based instance segmentation and tracking pipeline. Agent 3 performs mask-guided conditional generation to repair the predictor conditioning state, implemented with MagicQuil~\cite{magicquill} in our system.

The framework follows an event-driven control policy. When fresh uplink frames are available, the edge updates the conditioning state using the received observations and performs prediction only when necessary to bridge short gaps. When expected frames are missing, Agent 1 rolls out future frames autoregressively to preserve low-latency visual continuity. If a novel object appears during such a missing-frame interval, Agent 2 transmits a compact mask event to the edge, and Agent 3 uses this signal to repair the conditioning state used by Agent 1. Prediction then resumes from the repaired state. In this way, the three agents play complementary roles: Agent 1 preserves continuity, Agent 2 supplies sparse semantic evidence, and Agent 3 injects that evidence back into the predictor state without requiring full-frame retransmission.

\subsection{Prediction Mechanism}
Let $c_t=\{X_{t-1},X_t\}$ denote the conditioning state maintained at the edge at prediction time $t$. More generally, the framework assumes a fixed-length conditioning context required by the underlying predictor; in our implementation with MCVD, this context consists of two frames. Given $c_t$, the Continuous Prediction Agent $A_1$ generates a horizon-$T$ future segment autoregressively:
\begin{equation}
\hat{X}_{t+i}=f_{\text{MCVD}}(c_t,\hat{X}_{t+1:t+i-1}), \quad i=1,\dots,T .
\end{equation}
This horizon-based formulation matches the inference interface of the predictor, while the generated segment is treated as a provisional continuation that can be replaced whenever fresher observations arrive.

The role of $A_1$ is to maintain a continuously available visual stream under incomplete frame delivery. Whenever new uplink frames are received, the edge refreshes the conditioning state using the latest ground-truth observations. When the incoming burst is incomplete, $A_1$ immediately predicts the missing suffix from the most recent valid conditioning state. Thus, the predictor is not used as a standalone forecasting module, but as an online continuity mechanism whose state is repeatedly updated by the frame-arrival process.

We formalize this behavior as a sliding-window state transition. When a new ground-truth frame arrives, it replaces the oldest element in the conditioning state. When no new observation is available, the predictor advances using its own generated outputs. For the two-frame case used in our experiments, the update rule is
\begin{equation}
c_{t+1}=
\begin{cases}
\{X_t,X_{t+1}\}, & \text{if ground-truth } X_{t+1} \text{ arrives},\\
\{\hat{X}_t,\hat{X}_{t+1}\}, & \text{otherwise}.
\end{cases}
\end{equation}
This recursive update is identical to the prediction-only baseline during ordinary outages. Our contribution is not to modify the predictor itself, but to intervene on its conditioning state when new semantic evidence becomes available, so that subsequent rollouts remain both temporally smooth and semantically updated.

\subsection{Novel-Object Trigger Agent}
The Novel-Object Trigger Agent $A_2$ runs on the vehicle and monitors the live camera stream for newly appeared instances. We use a YOLO-based instance segmentation model with tracking to maintain object identities across frames. A novel object is defined as an instance whose track ID is absent from the recent history window. When such an event is detected at time $t^*$, $A_2$ sends a compact binary mask $M_{t^*}$ (and its timestamp) to the edge. This event-driven signaling is bandwidth-lightweight compared with transmitting full frames and is robust to uplink constraints.

Formally, let $\mathcal{I}_t$ be the set of tracked instance IDs at time $t$, and let $\mathcal{H}_t=\bigcup_{j=1}^{K}\mathcal{I}_{t-j}$ denote the ID history over the last $K$ frames. A novel object is detected when $\mathcal{I}_t \setminus \mathcal{H}_t \neq \emptyset$. The trigger agent then generates a mask $M_t$ by unioning the pixel regions of all newly appeared instances:
\begin{equation}
M_t=\bigcup_{o \in \mathcal{I}_t \setminus \mathcal{H}_t}\mathbb{1}_{\text{mask}}(o).
\end{equation}
Only the binary mask and a timestamp are transmitted. This design keeps the uplink payload small (on the order of kilobytes) and makes triggering feasible even under severe bandwidth constraints.

If multiple novel objects appear within a short interval, their masks are merged into a single trigger payload. If no new object is detected, the agent remains silent, thereby avoiding unnecessary communication. This selective signaling is critical for preserving uplink capacity while still enabling semantic correction when it matters.

The trigger message is buffered at the edge and aligned to the closest predicted-frame timestamp. This alignment ensures that the reconditioning agent operates on a temporally consistent predictor state, reducing visual discontinuities after correction.

\subsection{Conditional Frame Generation}
The Conditional Frame Reconditioning Agent $A_3$ is activated only when a novel-object trigger is received during a prediction interval. Its purpose is not to replace the displayed video stream directly, but to repair the internal conditioning state of the predictor so that the missing object can be propagated by subsequent rollouts. Let $t^*$ denote the trigger time, and let $\tau$ be the closest predicted timestamp available at the edge. In our two-frame instantiation, the reconditioning target is the aligned conditioning pair $\{\hat{X}_{\tau-1},\hat{X}_{\tau}\}$ together with the transmitted mask $M_{t^*}$.

Using MagicQuill, $A_3$ performs mask-guided conditional generation on the aligned predicted frames to obtain repaired conditioning frames:
\begin{equation}
\tilde{X}_{\tau-1}=f_{\text{MQ}}(\hat{X}_{\tau-1},M_{t^*}), \qquad
\tilde{X}_{\tau}=f_{\text{MQ}}(\hat{X}_{\tau},M_{t^*}).
\end{equation}
More generally, the number of frames to be reconditioned follows the conditioning length required by Agent 1. In our implementation, because MCVD uses a two-frame context, we repair two consecutive predicted frames around the aligned trigger time. This keeps the repaired state compatible with the predictor interface and preserves short-range temporal coherence before autoregressive rollout resumes.

The repaired conditioning state is then used to reinitialize the predictor:
\begin{equation}
\tilde{c}_{\tau}=\{\tilde{X}_{\tau-1},\tilde{X}_{\tau}\}.
\end{equation}
Agent 1 subsequently continues prediction from the repaired state as
\begin{equation}
\hat{X}_{\tau+i}=f_{\text{MCVD}}(\tilde{c}_{\tau},\hat{X}_{\tau+1:\tau+i-1}), \quad i=1,\dots,T .
\end{equation}
A key design choice is that the repaired frames are injected into the predictor state rather than directly substituted into the output stream. Direct display replacement would create a mismatch between the visible frame and the latent temporal state from which future predictions are generated, often causing abrupt drift immediately after correction. By instead repairing the conditioning state, the framework allows $A_1$ to absorb the newly introduced object and propagate it forward in a temporally coherent manner. As a result, the additional system overhead is limited to sparse mask transmission and occasional reconditioning inference, while preserving the bandwidth advantages of prediction-based teleoperation.

\section{Evaluation and Discussion}
\label{eval}

\subsection{Experimental Setup}
\noindent \textbf{Dataset and Evaluation Scenarios}
We evaluate our approach on video sequences from the Cityscapes dataset~\cite{cordts2016cityscapes}, which provides real-world urban driving scenes with naturally occurring dynamic objects. The videos include spontaneous object appearances (e.g., pedestrians or vehicles entering the field of view), enabling evaluation under open-world conditions without synthetic insertion. We evaluate multiple sequences containing diverse novel-object events to ensure result robustness.

For each sequence, the first two frames are used as conditioning frames to initialize the predictive model. Subsequent frames evolve according to the original video dynamics. To ensure fair comparison across methods, frame counts and conditioning windows are controlled consistently when computing reconstruction metrics.

\noindent \textbf{Communication Simulation.}
To emulate realistic teleoperation conditions, we replay uplink throughput traces collected from real-world 5G deployments, as shown in Figure~\ref{fig:uplink_analysis}. The trace dataset consists of approximately 18 hours of uplink measurements recorded at 100 ms intervals under diverse mobility conditions, including handover events and bandwidth fluctuations.

During simulated outage periods (uplink $\leq 5$ Mbps), full-frame transmission is constrained by measured instantaneous bandwidth, and the edge must rely on predictive streaming to maintain visual continuity.

\noindent \textbf{Novel Object Events and Metrics.}
Novel-object events are identified using a vehicle-side YOLO-based segmentation and tracking module. When a previously unseen object instance enters the scene, a detection trigger is generated and transmitted to the edge as a lightweight mask signal. For evaluation, the timestamp of the first ground-truth appearance of the object in the video is recorded as the reference time.

To quantify Novel-Object Blindness, we define Scene Recovery Time (SRT) as the interval between (i) the ground-truth first appearance of a novel object and (ii) the first frame in the multi-agent output stream where the object becomes visible. SRT is measured within each 20-frame segment and directly captures semantic recovery behavior during communication gaps. In addition to SRT, we report standard video quality metrics, including FVD, LPIPS, MSE, PSNR, and SSIM, to assess overall perceptual fidelity.

All experiments are conducted on a single NVIDIA RTX A6000 GPU to emulate edge-side inference.

\begin{figure}[t]
  \centering
  \includegraphics[width=0.95\linewidth]{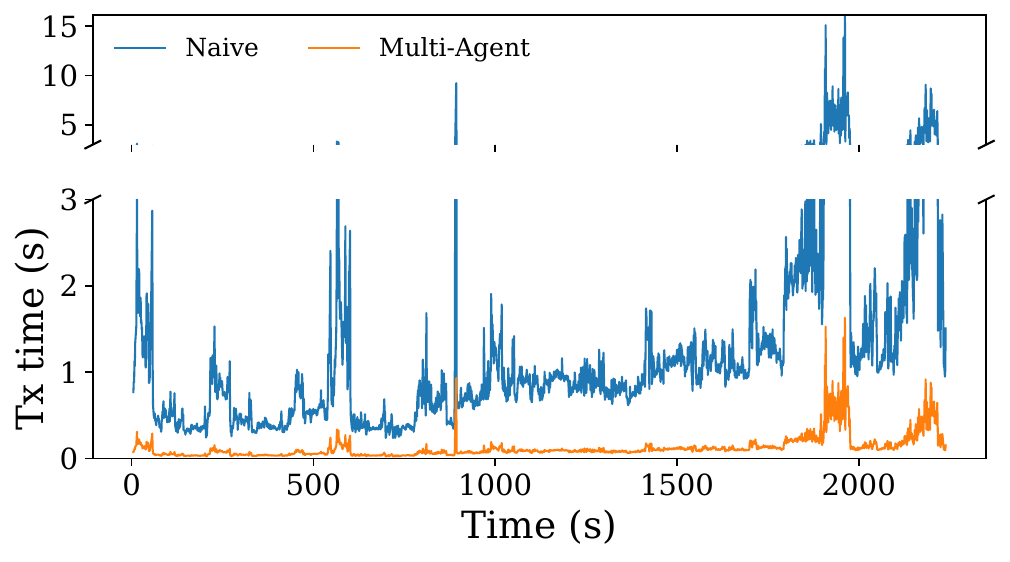}

   \caption{Instantaneous transmission time under the OPX uplink trace for the Naive approach and our proposed method.}
   \label{fig:tx_time}
\end{figure}

\subsection{Semantic Recovery through Conditional Frame Generation}
\label{sec: nob_communication_gap}
\begin{figure*}[t]
    \centering
    \includegraphics[width=0.95\linewidth]{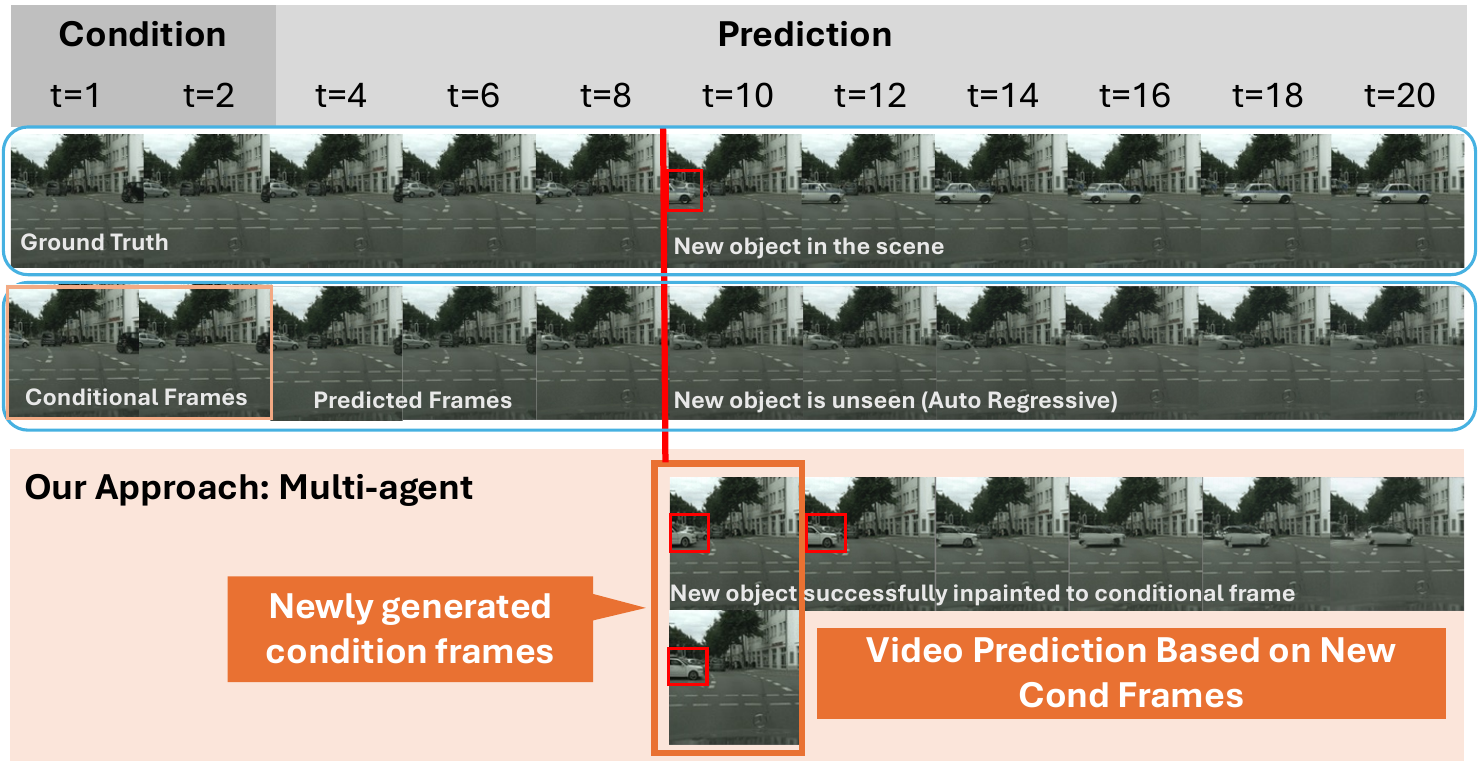}
    \caption{Qualitative comparison among ground truth, prediction-only streaming, and our approach.}
    
    \label{fig:comparison}
    \vspace{-1.5em}
\end{figure*}

We evaluate how the proposed multi-agent framework restores semantic consistency when novel objects emerge during uplink interruptions. Figure~\ref{fig:comparison} provides a qualitative comparison among ground truth, prediction-only streaming, and our multi-agent framework. The first row shows the ground-truth sequence, the second row corresponds to prediction-only streaming (Agent 1), and the third row shows our multi-agent framework (prediction + mask-guided reconditioning).

The predictive model is initialized with the first two frames (t=1 and t=2) as conditioning inputs. Upon receiving these frames, the edge predictor autoregressively generates future frames (t=3--12) in the absence of additional observations. At t=9, a previously unseen white vehicle enters the scene from the lateral direction while the ego vehicle continues moving forward. In the ground-truth sequence, this crossing vehicle introduces a potential collision risk if the ego vehicle does not decelerate.

However, in the prediction-only stream, the white vehicle is absent despite continued temporal smoothness of the generated frames. Since the object was not present in the initial conditioning window, the autoregressive predictor fails to incorporate it into the evolving scene, resulting in a semantically incomplete video stream. In teleoperation, an operator observing such predicted frames would not be aware of the crossing vehicle and may fail to adjust speed accordingly.

In contrast, our multi-agent framework receives a mask of the newly detected white vehicle from the vehicle-side perception module. A lightweight inpainting agent generates an updated conditioning frame that incorporates the new object, after which the predictive agent is reconditioned and resumes autoregressive generation. As shown in the third row of Figure~\ref{fig:comparison}, the crossing vehicle is correctly introduced into subsequent frames, aligning the predicted stream with scene evolution.

To quantify this effect, Table~\ref{tab:srt} reports Scene Recovery Time (SRT) statistics under novel-object appearance. SRT measures the interval between the ground-truth first appearance of a novel object and its first visible reconstruction in the output stream. Prediction-only streaming exhibits a mean SRT of 500 ms, reflecting prolonged semantic blindness during communication gaps. Our multi-agent framework reduces the mean SRT to 275 ms, with substantially lower minimum recovery latency. These results confirm that the proposed reconditioning mechanism effectively shortens NOB duration.

We further evaluate overall perceptual fidelity in Table~\ref{tab:main_results}. The proposed approach achieves a substantial FVD improvement, indicating closer alignment with ground-truth video dynamics at the sequence level. Pixel-level metrics such as PSNR, SSIM, and MSE show only marginal differences relative to prediction-only streaming. These small variations arise from corrective object insertion during reconditioning but do not materially degrade visual continuity. Overall, the results show that terminating Novel-Object Blindness improves semantic consistency without significantly compromising low-level reconstruction quality.

Together, the qualitative and quantitative results demonstrate that the proposed multi-agent reconditioning framework restores semantic consistency more effectively than prediction-only streaming in open-world settings, with minimal impact on overall perceptual fidelity. We note that SRT focuses on semantic recovery within 20-frame segments and does not directly represent end-to-end operator response latency.

\begin{table}[t]
\centering
\caption{Scene Recovery Time (SRT) statistics under novel object appearance (FPS=20), measured within each 20-frame segment in the output stream.}
\label{tab:srt}
\begin{tabular}{lcc}
\toprule
Method & Mean SRT (ms) & Min--Max (ms) \\
\midrule
Pred-only & 500 & 500--900 \\
Multi-Agent & \textbf{275} & 50--500 \\
\bottomrule
\end{tabular}
\end{table}



\begin{table}[t]
  \caption{Quantitative comparison on video prediction.
  $\Delta$ denotes (Ours - Baseline).}
  \label{tab:main_results}
  \centering
  \begin{tabular}{@{}l S S S@{}}
    \toprule
    Metric & {Pred-only} & {Ours} & {$\Delta$} \\
    \midrule
    FVD $\downarrow$   & 437.56 & 178.75 & -258.81 \\
    LPIPS $\downarrow$ & 0.0579 & 0.0724 & 0.0145 \\
    MSE $\downarrow$   & 0.00272 & 0.00347 & 0.00075 \\
    PSNR $\uparrow$    & 25.69 & 24.63 & -1.06 \\
    SSIM $\uparrow$    & 0.9207 & 0.8936 & -0.0270 \\
    \bottomrule
  \end{tabular}
  \vspace{-1.5em}
\end{table}

\subsection{Runtime and System Overhead}
We finally analyze the runtime overhead of the proposed framework under communication gaps. Table~\ref{tab:latency_breakdown} presents a latency breakdown for recovering a 20-frame outage under a 5 Mbps uplink constraint.

Under the naïve streaming strategy, all 20 frames must be transmitted directly, resulting in a total latency of 3280 ms. Prediction-only streaming reduces this to 1228 ms by transmitting only two conditioning frames and generating the remaining frames at the edge. Our multi-agent framework incurs a total latency of 1441 ms, which includes additional costs for mask generation, mask transmission, and lightweight inpainting.

Importantly, the added overhead relative to prediction-only streaming is modest while remaining substantially lower than the naïve full-frame transmission baseline. The additional latency corresponds to the corrective reconditioning step that enables termination of Novel-Object Blindness. Given the significant Scene Recovery Time reduction demonstrated in Section~\ref{sec: nob_communication_gap}, this overhead represents a favorable trade-off between semantic reliability and transmission efficiency.

\subsection{Communication-Constrained Deployment Analysis}
We also analyze whether the proposed multi-agent pipeline remains deployable under realistic uplink perturbations. Figure~\ref{fig:tx_time} compares the instantaneous transmission time of naïve full-frame streaming and our multi-agent framework under the OPX uplink trace.
Under the naïve strategy, each frame is transmitted directly over the uplink. As bandwidth fluctuates, transmission time increases sharply in low-throughput intervals, producing pronounced latency spikes that correspond to outage periods.

In contrast, our framework decouples stream continuity from instantaneous uplink bandwidth. During perturbations, only conditioning frames and lightweight mask signals are transmitted, while intermediate future frames are generated at the edge. As shown in Figure~\ref{fig:tx_time}, this design substantially suppresses transmission-time bursts relative to naïve streaming.

These results indicate that the proposed multi-agent reconditioning pipeline can sustain low-latency operation under communication constraints, supporting practical edge deployment without continuous high-bandwidth connectivity.

Overall, the results indicate that the proposed multi-agent design achieves timely semantic recovery with marginal system overhead, making it suitable for real-time edge deployment under realistic 5G constraints.

\begin{table}[t]
\centering
\small
\caption{Latency breakdown for recovering a 20-frame gap (uplink=5 Mbps, image=100 KB/frame, mask=2 KB).}
\label{tab:latency_breakdown}

\begin{tabular}{p{1.15cm} p{2.95cm} r r}
\toprule
Method & Comp. & Time (ms) & Total (ms) \\
\midrule

Naive
& Transmit 20 frames & 3280 & \textbf{3280} \\
\midrule

Pred-only
& Transmit 2 cond frames & 328 & \textbf{1228} \\
& Predict 18 frames & 900 &  \\
\midrule

Multi-Agent
& Transmit 2 cond frames & 328 & \textbf{1441} \\
& Generate mask & 20 &  \\
& Transmit mask & 3 &  \\
& Inpaint 2 frames & 290 &  \\
& Predict 16 frames & 800 &  \\
\bottomrule
\end{tabular}
\vspace{-2.0em}
\end{table}







\section{Conclusion}

This paper studies semantic reliability in edge-assisted predictive video streaming for teleoperation under realistic 5G uplink perturbations. We show that although autoregressive prediction preserves short-term visual continuity during communication gaps, it suffers from a structural failure mode under open-world scene evolution: newly appeared objects outside the conditioning history may be systematically omitted. We formalize this phenomenon as Novel-Object Blindness (NOB) and address it with a multi-agent reconditioning framework that couples continuous edge prediction with mask-guided lightweight inpainting.

Experiments under real 5G uplink traces show that the proposed design improves semantic recovery under communication gaps while preserving overall perceptual quality and real-time feasibility. In particular, Scene Recovery Time is substantially reduced compared with prediction-only streaming, and system overhead remains modest relative to naive full-frame transmission. These results indicate that semantic consistency can be improved without sacrificing practical deployability in edge teleoperation settings.
\section*{Acknowledgments}
The research was supported in part by NSF awards CNS-2321531, DMS-2436333 and ITE-2453815 and MnDOT/LRRB grant.
{
    \small
    \bibliographystyle{ieeenat_fullname}
    \bibliography{main}
}


\end{document}